\input harvmac

\def \s {\sigma}
\def\t {\tau}

\def \ha {\half}
\def \ov {\over}

\def \a {\alpha}

\def \lr { \lref}

\def\del {\partial }

\def\l{\lambda}

\def \td {\tilde }
\def \k {\kappa}

\def\rs { ${\bf R}^{10}\times S^1 $}
\def\rss { ${\bf R}^{9}\times S^1 \times S^1 $}

\gdef \jnl#1, #2, #3, 1#4#5#6{ { #1~}{ #2} (1#4#5#6) #3}

\def\np {  Nucl. Phys. }
\def \pl { Phys. Lett. }

\def \ijmp {Int. J. Mod. Phys. }
\def \cmp {Commun. Math. Phys. }

\lr \berg { E. Bergshoeff,  E. Sezgin and P.K. Townsend,
\pl {\bf B189} (1987) 75;
 Ann. Phys. {\bf 185} (1988) 330.} 

\lr\dewit { B. de Wit, M. L\" uscher and H. Nicolai, 
Nucl. Phys. {\bf B320} (1989) 135.}

\lr\dewitt { B. de Wit, J. Hoppe and H. Nicolai, 
Nucl. Phys. {\bf B305} [FS 23] (1988) 545.}

\lr\kaku{ M. Kaku, 
hep-th/9606057; hep-th/9607111.}
\lr\bars{ I.  Bars, Nucl. Phys. {\bf B343} (1990) 398.}

\lr\town{  P.K. Townsend, Phys. Lett. {\bf B350} (1995) 184.}
\lr\witt{ E. Witten, \np {\bf B443} (1995) 85 .}

\lr\duste{M.J. Duff and K.S. Stelle, \pl {\bf B253} (1991) 113.}

\lr \bps {I. Bars, C.N. Pope and E. Sezgin, Phys. Lett. {\bf 198} (1987) 455.}

\lr\hora {  P. Ho\v rava and E. Witten, Nucl. Phys. {\bf B460} (1996) 506;
E. Witten, \np {\bf B463} (1996) 383.}

\lr\dlp { J. Dai, R. Leigh and J. Polchinski, Mod. Phys. Lett. {\bf A4} (1989) 2073.}

\lr\simo {B. Simon, Ann. Phys. {\bf 146} (1983) 209.}

\lr\rutse{ J.G. Russo and A.A. Tseytlin, hep-th/9611047. }

\lr\simon {B. Simon, J. Func. Anal. {\bf 53} (1983) 84.}

\lr\hoppe{J. Hoppe, \ijmp {\bf A4} (1980) 5235.} 

\lr\toro{D. Fairlie, P. Fletcher and C. Zachos, \pl {\bf B218} (1989) 203;

J. Hoppe, \ijmp {\bf A4} (1989) 5235;

B. de Wit, U. Marquard and H. Nicolai, \cmp {\bf 128} (1990) 39.}

\lr\jevicki{ A. Jevicki, hep-th/9607187.}

\lr\schw{For a review, see J. Schwarz, hep-th/9607201.}

\lr\toww {P. Townsend, \pl {\bf B373} (1996) 68; C. Schmidhuber,
hep-th/9601003; A.~Tseytlin, hep-th/9602064.}


\lr\duff{ M. Duff, T. Inami, C. Pope, E. Sezgin and K. Stelle,
\np {\bf B297} (1988) 515.}
\lr\russo { J.G. Russo, \pl {\bf B} (1997) , hep-th/9609043. }

\lr\dduff{ M. Duff, P. Howe, T. Inami  and K. Stelle, \pl {\bf B191}
(1987) 70.}
\lr\bst{E. Bergshoeff,  E. Sezgin and Y. Tanii, \np {\bf B298}
(1988) 187.}

\baselineskip8pt
\Title{\vbox
{\baselineskip 6pt{\hbox{CERN-TH/96-275}} {\hbox{hep-th/9610018}} {\hbox{
   }}} }
{\vbox{\centerline {Supermembrane dynamics from multiple}
\centerline        {interacting strings }
}}
 
\vskip -20 true pt

\centerline  { J.G. Russo
\footnote {$^*$} {e-mail: jrusso@vxcern.cern.ch}
 }

 \smallskip \bigskip
 
\centerline{\it  Theory Division, CERN}
\smallskip

\centerline{\it  CH-1211  Geneva 23, Switzerland}

\bigskip\bigskip\bigskip
\centerline {\bf Abstract}
\bigskip

The supermembrane theory on \rs\ 
is investigated, for membranes that wrap once around the compact
dimension. The Hamiltonian can be organized  as describing 
 $N_s$  interacting strings,
 the exact supermembrane corresponding to $N_s\to \infty $.
The zero-mode part of $N_s-1$ strings 
turn out to be precisely the modes 
which are  responsible of instabilities.
For sufficiently large compactification radius $R_0$,  interactions 
are negligible and the lowest-energy excitations are described by
a set of harmonic oscillators.
We compute  the physical spectrum to leading order, which becomes exact 
in the limit $ g^2 \to \infty $, where $g^2\equiv 4\pi ^2 T_3 R_0^3$ and $T_3$ 
is the membrane tension.
As the radius is decreased,  more strings become strongly
interacting and their oscillation modes get frozen. 
In the zero-radius limit, the spectrum is constituted of the 
type IIA superstring spectrum, plus an infinite number of extra states
associated with flat directions of the quartic potential.

\medskip
\baselineskip8pt
\noindent

\Date {October 1996 }

\noblackbox
\baselineskip 17pt plus 2pt minus 2pt

\vfill\eject
\newsec{ Introduction}
 
It is expected that
supermembrane theories \berg\ will play a central role in
a new formulation of string theory   
that should incorporate features of the strong coupling dynamics, which
is now referred to as  M-theory \schw .
In the  eleven-dimensional supermembrane theory, the existence of a supersymmetric
 ground state, together with the assumption of an energy gap to the first excited state,
would be  sufficient to guarantee that the low-energy limit of the theory will 
indeed be given by 11D supergravity, believed to govern the strong 
coupling limit of type IIA superstring theory \refs {\town , \witt }.
Moreover, 
supermembrane theory is
known to be related to type IIA superstring theory by means of a procedure
called double-dimensional reduction \dduff .
The idea is to consider the target space \rs , fix a gauge where
the compact coordinate is proportional to one of the world-volume coordinates, 
say $\rho $, and  --in the  spirit of Kaluza-Klein reduction-- 
assume that for small radius $R_0$ the relevant modes of the 
non-compact coordinates will be those independent of $\rho $.
In this work we will incorporate the dynamics of  other modes
in a systematic way.
 
Strictly speaking, the statement that type IIA superstring theory
is obtained from the supermembrane  at small radius 
still remains to be proved. The actual
statement of ref.~\dduff\ is that type IIA superstring theory
is obtained under the {\it ad hoc} assumption
that the non-compact coordinates are $\rho $-independent.
  As we will see, the problem is that there are extra 
states, associated with  the
instabilities of the quantum supermembrane on ${\bf R}^{11}$,  which do not
decouple as $R_0\to 0$.
This will be clear in the present formulation, where  the supermembrane will be 
organized as a system of   interacting strings.
In this language, the  extra states include
the states associated with instabilities on \rs ,  
which are constructed by using the zero-mode part of $N_s-1$ strings.
 In order for  {\it standard} supermembrane theory 
to have a discrete mass spectrum,
 these zero modes have to be removed by hand from the original Hamiltonian.
Since the action of \berg\ is not renormalizable,
 it is conceivable that in a correct description, appropriate
to $R_0$ less than the Planck length,
the unwanted states could decouple.
Here we will keep all the modes of supermembrane theory, except
at the end of sect. 6, where  we will consider the truncated Hamiltonian
with the zero modes of  $N_s-1$ strings  removed.
It describes a stable system, with a calculable discrete spectrum
at large radius.

Another possibility \refs{\dewit, \russo } 
  is that  quantum effects could modify the asymptotic zero-point energy in certain 
  directions, in such a way that all modes remain confined, and as a result the Hamiltonian
 spectrum would be discrete. The problem is how to do this maintaining
 supersymmetry.
The mechanism explored in ref. \russo\  uses a target-space with a boundary,
where the membrane wave function is required to vanish.
This requirement indeed removes the instabilities, but,
on the face of it,  it seems to also preclude the existence of 
 massless states in the spectrum. Unfortunately,
the general ground-state wave-functional is  unknown, so 
 it is not clear whether there could remain some eigenstate of zero energy
  satisfying the boundary condition.

In standard Kaluza-Klein theory, the extra dimension emerges 
in scattering processes involving energies greater than the inverse compactification
radius, $1/R_0$, i.e. when  Kaluza-Klein
modes  can be excited. Processes involving
energies which are much smaller than $1/R_0$,
can be effectively described in terms of a $D-1$ dimensional theory.
In the present case, one would like to see whether for energies much
lower than $1/R_0$ the theory can be effectively described in terms
of a string theory, i.e. whether the only 
states that can be excited at $E\ll 1/R_0$ correspond
to oscillation modes of a string.
 In what follows, when discussing the regime $R_0^2\ll \a' $, by ``low-energy" we mean
$E^2\leq O(1/\a')\ll 1/R_0^2$. This is the regime where we would
expect to have a string theory description, with all the  extra 
quantum states of the
supermembrane frozen.

The dynamics dictated by the membrane Hamiltonian
 will be investigated in sects.~4 and ~5. 
Although the full Hamiltonian is highly non-linear, 
the qualitative behavior of the system can be followed for all radii and,
remarkably, exact statements about the relevant excitations of the system
can be done in the extreme limits $R_0=\infty $ and $R_0=0$;
from the ten-dimensional viewpoint, these correspond respectively 
to infinite and zero string coupling 
$g^2\equiv 4\pi ^2T_3R_0^3$.
In the $R_0=\infty $ limit, with $T \equiv 2\pi R_0 T_3$ fixed,
the Hamiltonian describes a system of decoupled harmonic oscillators.
For finite but large $R_0 $, quantum states with 
 $\a' M^2\ll O(g^2 )$ will describe harmonic motion, while
 those with $\a ' M^2 > O(g^2  )$ will be affected by the 
 full non-linearities of the supermembrane theory.
 In the limit $R_0\to 0$, the quantum states 
 with  non-trivial oscillator content in the compact direction
will have $M^2\sim 1/R_0^2$, 
  and  they get frozen: 
it takes an infinite energy to excite them. 

\newsec{ Supermembrane Hamiltonian and instabilities }

\subsec{  Generalities }

We will consider membranes of toroidal topologies. A 
 complete set of functions
on the torus is given by
\def\n{ {\bf n} }
\def\m{ {\bf m} }
\eqn\base{
 Y_{\n }(\s ,\rho )= e^{in_1\s +in_2\rho }\  , \ \ \   \n =(n_1,n_2)\ ,\ \  
\s , \rho \in [0, 2\pi )\ .
}
The Lie bracket of $ Y_{ \n }(\s )$ corresponds to the structure
constants of the area-preserving diffeomorphisms of the torus,
\eqn\lieb{
\{ Y_{ \n }, Y_{ \n '} \} = f_{\n \n ' \m } Y^\m =
- (\n \times \n ' ) Y_ {\n +\n '} \ ,
}
where 
$$
\{ X, Y \} = \partial _ {\s } X  
\partial _ {\rho } Y - \partial _ {\rho } X  
\partial _ {\s } Y \ ,
\ \ \ \ \n \times \n ' = n_1 n_2' -n_2 n_1' \ .
$$
There is a truncated version of this infinite dimensional algebra,
which approximates eq.~\lieb\ in the limit $N\to \infty $:
\eqn\rlb{
 [ J_{ \n }, J_{ \n '} ]= g_{\n \n ' \m } J^\m=
-{N\ov 2\pi }\sin { 2\pi \ov N} (\n \times \n ' ) J_ {\n +\n '} \ .
}
One can restrict to a fundamental lattice defined by 
$n_1, n_2=0,..., N-1$, with the exception of the origin $n_1=n_2=0$, which label
$N^2-1$ generators; these can be shown to span 
the algebra of $SU(N)$ \toro .
Note that the Cartan subalgebra can be taken to be generated by 
$J_{ \n }$, with $n_1=0$, $n_2=1,..., N-1$.

Taking
$X^{9}$ as one of the light-cone variables,
$X^\pm= {X^0\pm X^9 \ov \sqrt {2} }\ $,
the  light-cone Hamiltonian of the supermembrane is given by 
\refs{\bst , \dewitt}
$$
H=H_{\rm B} + H_{\rm F} \ ,
$$
\eqn\lcgh{
H_{\rm B} =  
2\pi ^2 \int d  \sigma d\rho \left [
 P _ a   ^ 2 +
{ T_3^2\over 2 }
( \{ X ^ a, X ^ b \} ) ^ 2\right ] \ ,
}
\eqn\lcgh{
H_{\rm F} =- T_3p_0^+  \int d  \sigma d\rho \ 
 \bar \theta \Gamma ^ a \{ X ^ a  , \theta \} \ ,\ \ \ \ a=1,2,...,8,10\ .
} 
Here $\theta $ are real $SO(9)$ spinors $\theta _\a $, $\a =1,...,16 $.
The mass square operator 
$M^2=2p^+_0p^-_0-(p_0^{a})^2$ is given by $M^2=2 H-(p_0^{a})^2$.
$T_3$ is the membrane tension, $[T_3]={\rm cm}^{-3}$.

The Hamiltonian has a residual gauge invariance, corresponding to the
symmetry \lieb\ of area-preserving diffeomorphisms of the torus.
Let us choose $Y_{(0,n_2)}$ as the generators of the Cartan subalgebra
$K$. We expand
\eqn\cartt{
X^ a(\s,\rho  )= \sum _\n  X^a_\n Y_{ \n }(\s ,\rho )\ 
=X_K^a(\s,\rho  ) + \tilde X^a (\s,\rho  ) \ ,
}
$$
X^a_K(\s,\rho  )=\sum _{\n \in K} X^a_\n Y_{ \n }(\s ,\rho )\ ,\ \ \ \ 
\tilde X^a(\s,\rho  )  =\sum _{\n , n_1\neq 0} X^a_\n Y_{ \n }(\s ,\rho ) \ .
$$
The area-preserving gauge symmetry can be fixed by setting 
$\tilde X^{10} (\s,\rho  )=0$ \dewit .

\subsec{  Instabilities in the presence of compact dimensions}

Consider the target space \rss , and expand the single-valued 
membrane coordinates
\eqn\xmu{
X^a (\s,\rho  )= \sum _\n  X^a_\n Y_{ \n }(\s ,\rho )\ ,\ \ \ \  a=1,...,8\ .
}
Let the 
coordinates $X^9, X^{10}$ have periods $2\pi R_9$ and $2\pi R_{10}$, respectively.
Now we perform a semiclassical expansion around a classical
 configuration with non-trivial winding numbers,
\eqn\xonc{
X^{9} (\s ,\rho  ) = w_1 R_9\s  +w_2 R _9\rho + \sum _\n  X^{9}_\n Y_{ \n }(\s ,\rho  )
\ ,
}
\eqn\xonc{
X^{10} (\s ,\rho  ) = v_1 R_{10}\s  +v_2 R _{10}\rho + 
\sum _\n  X^{10}_\n Y_{ \n }(\s ,\rho)
\ .
}
The small-oscillation spectrum in the case  $w_2=v_1=0$ was derived in
 ref. \duff .
As in sect. 2.1, the light-cone gauge can be used to remove 
the single-valued part of $X^{9}$, and the residual symmetry of 
area-preserving diffeomorphisms can be fixed as before 
by constraining $X^{10}$ to live in the Cartan subspace of the 
algebra \lieb , that is
\eqn\xonc{
X^{9} (\s ,\rho  ) = w_1 R_9\s  +w_2 R _9\rho +X^9_{(0,0)} \ ,
}
\eqn\xonc{
X^{10} (\s ,\rho  ) = v_1 R_{10}\s  +v_2 R _{10}\rho + 
X^{10}_K (\s ,\rho )\ ,
\ \ \ \ X^{10}_K (\s ,\rho )=\sum _{\n \in K} X^{10}_\n Y_{ \n }(\s ,\rho  )
 \ .
}
The winding number that counts how many times the toroidal membrane
is wrapped around the target-space torus is given by
\eqn\wiwi{
w_0={1\ov 4\pi^2 R_9 R_{10} }\int d\s d\rho \ \{ X^9, X^{10}\} 
= w_1 v_2-w_2v_1 \ .
}
For $w_0\neq 0$ the membrane is expected to be stable for topological reasons.
Let us see how this works using  the Hamiltonian formulation.
Because of the winding contributions, there will now be  new terms in   
the Hamiltonian, which   will in turn be responsible for the confinement  
of all membrane modes. In particular, the bosonic interaction term is now 
given by
\eqn\aaaa{
\pi ^2 T_3^2\int d  \sigma d\rho 
( \{ X ^ \mu, X ^ \nu \} ) ^ 2 = \pi ^2 T_3^2\int d  \sigma d\rho 
 ( \{ X ^ a, X ^ b \} ) ^ 2  
}
$$
+4\pi ^2 T_3^2R_{10}
\int d\sigma d\rho  \ \{ X^{10} ,  X^i \} 
(v_2\del_\s X^i-v_1\del_\rho X^i)
  $$
$$
+\  8\pi^4 T_3^2 \bigg( \sum_\n
\big[ R_9^2 ({\bf w}\times \n )^2  X_\n^a X_{-\n}^a +
R_{10}^2 ({\bf v}\times \n )^2  X_\n^a X_{-\n}^a \big]+ 
R_9^2 R_{10}^2 ({\bf w}\times {\bf v} )^2 \bigg)
 \ . 
$$
Here $\mu ,\nu =1,...,10$ and $a,b=1,...,8,10$, where the replacement
 $X^{10}\to X^{10}_K$ is understood on the right-hand side. 
Let us choose $Y_{\bf v}$ as one of the Cartan generators.
 Instabilities due to flat directions
in the full Hamiltonian can only occur along the Cartan directions \dewit ;
 for all $\n \in K$ we will have ${\bf v}\times \n =0$ .
Let us take for example $v_1=0, v_2=1$, so that the Cartan subalgebra will be generated by $Y_{(0, n_2)}$. Provided we choose ${\bf w} $
outside the Cartan subspace, 
that is ${\bf w}\times {\bf v}= w_1\neq 0$,
 then the  term in eq. \aaaa\ containing $({\bf w}\times \n )^2   $
will be non-zero for all 
Cartan directions $\n =(0,n_2)$ and, as a result, all modes will
be confined.
 Thus, as expected, the Hamiltonian will  have a discrete spectrum
precisely when the membrane is topologically protected, i.e. 
$w_0={\bf w}\times {\bf v}\neq 0$.
Note that this is true in the large radius regime where the
semiclassical approximation is justified (for small radius, 
world-volume instanton effects may lead to  instabilities).

If the target space is given by \rs , then the stability of the 
membrane cannot be topologically
protected. It is clear from the above analysis that there will be flat directions: there is only one term 
$\ha R_{10}^2({\bf v}\times \n )^2  X_\n^a X_{-\n}^a $, which vanishes
for certain $\n $. In particular, taking ${\bf v} =(0,1)$, 
wave packets of the form $\chi (X_{(0,n_2)}- V t) $, with $\chi $ of compact support, can escape to
infinity.

\newsec{Hamiltonian for multiple strings}

\subsec{Strings from membranes }
Let the target space be given by \rs ,
with $X^{10}$ representing the compact dimension, and 
 take
$X^{9}$ as one of the light-cone variables,
$$
X^\pm= {X^0\pm X^9 \ov \sqrt {2} }\ .
$$
Consider a configuration that wraps once around $S^1$,
and fix the symmetry of area-preserving diffeomorphisms
by setting
\eqn\uno{
X^{10}=R_0 \rho +X^{10}_K(\rho )\ ,
}
where $X^{10}_K(\rho )$ is a single-valued function that lives
in the Cartan subspace generated by $Y_{(0,m)}$,
$$
X^{10}_K(\rho )=\sum _{m} X_{(0,m)} ^{10} e^{im\rho }\ .
$$
Let us write 
$P^{10}(\s ,\rho )=P_K^{10} + \td P^{10}  $, where $P_K^{10}$ 
lives in the Cartan
subspace, and $\td P^{10}$ in the complement.
As usual, the constraints associated with the gauge symmetry
can be either imposed on the quantum states or they can be solved
at the classical level to eliminate one of the variables as an 
independent degree of freedom to be quantized. 
A description of the local and global 
constraints in the light-cone gauge can be found in the appendix.
The local constraints can be solved for $\td P^{10}$ in terms of
$X^{10}_K$ and the transverse coordinates and momenta $X^i , P^i$,
$i=1,...,8$ (see eqs.~(A.6),~(A.8)).

In the following, we will concentrate on the bosonic part. 
The incorporation
of fermions is straightforward and will be discussed in sect. 6.
Using eq. \uno , the Hamiltonian takes the form  
($f'(\s )\equiv\del _\s f(\s )$ throughout)
\eqn\hhh{
H_{\rm B}=H_0+ H_1 \ ,
}
\eqn\hhcc{
H_0=2\pi ^2 \int d  \sigma d\rho \left [
  P _ a   ^ 2 +
T_3^2 R_0^2 (X_i')^2 \right ] \ ,
}
\eqn\hhpp{
H_1=\pi ^2 T_3^2 \int d  \sigma d\rho 
\big[ ( \{ X ^ a, X ^ b \} ) ^ 2+ 4R_0 \del_\rho X^{10}(X_i')^2\big]\ ,
}
$$
a,b= 1,...,8,10\ ,\ \ \ \ \ \ \ \ i,j=1,...,8\ ,
$$
understanding $X^{10}\to X^{10}_K(\rho) ,\ \ 
P^{10}\to P^{10}_K(\rho)+\td P^{10}$, where $\td P^{10}$ is given by 
eq.~(A.6).

Let us now expand the transverse bosons in a complete basis of functions on $S^1$:
\eqn\bbss{
X^i(\s , \rho ,\tau )= \sum _m X_m^i(\s ,\tau  ) e^{im \rho } \ ,
\ \ \ P^i (\s ,\rho ,\tau )= {1\ov 2\pi }\sum _m P_m^i(\s ,\tau ) e^{im \rho } 
}
$$
[X^i_m(\s ), P^j_n(\s ' )]=i\delta _{m+n }\delta ^{ij}\delta (\s -\s ' )\ 
,\ \ \ \ X^\dagger_m= X_{-m}\ ,
\ \ \ P^\dagger_m= P_{-m}\ .
$$
Hermitian variables $Y_m^{(1,2)}$ can be introduced  through
$ X_{\pm m}=Y_m ^{(1)}\pm iY_m^{(2)}\ $, $m=1,2,... $, and similarly for
$P_m $. 

Inserting the expansions \bbss , the Hamiltonian \hhh\ 
reduces to the following expressions:
\eqn\aaa{
H_0=\pi T^2 \int d\s 
\sum _m \left [
 T ^{-2} P^a_{m} P^a_{-m} + X_{m}^{a\prime }{X_{-m}^{a\prime }} \right ]
\ ,\ \ \ \ T \equiv 2\pi R_0T_3\ , 
}
\eqn\bbb{
H_1= {\pi T^2\ov R_0^2} \int d\s 
\sum _{m,n,p} \big[
mp ({X_n^{a\prime }} X_p^a) ( X^{b\prime }_{-m-n-p} X_m^b) -
}
$$
np (X_n^a X_p^a) ( X^{b\prime }_{-m-n-p} X_m^{b\prime })
\big]\ 
$$
$$
+ \ {2\pi iT^2\ov R_0}\int d\s \sum _{m,n} \  m 
X^{10}_{(0,m)} X^{i\prime }_{n}X^{i\prime}_{-n-m}
$$
In this context, the usual double-dimensional reduction procedure \dduff\ corresponds
to dropping all modes $X_m^a(\s )$ with $m\neq 0$, 
and setting  the Kaluza-Klein momentum
$p^{10}=\int d\rho P^{10}(\rho)$ to zero,
so what remains is 
\eqn\sredd{
\a'  H_{\rm B}^{\rm red }={T \ov 2 }\int d\s 
 \left [ T ^{-2}P^i_{0} P^i_{0} +
{X_{0}^i}'{X_{0}^i}' \right ]  \ ,\ \ \ \ \a' \equiv (2\pi T )^{-1}\ .
}
 Upon the identification of $T $ with the standard string tension,
this is exactly the  string theory Hamiltonian.

World-volume time translations are generated by 
$\tilde H=\a'  ( H_0 + H_1)$. From the Hamilton equations, one 
obtains
$$
P_m^i=T  \del_\tau X^i_m \ .
$$
Regarding $H_1$ as a perturbation, the equations of motion of the
unperturbed Hamiltonian $H_0$ give
\eqn\eas{
\del_\s^2 X_m^i= \del_\tau ^2 X_m^i\ .
}
The solution satisfying the periodicity condition $X^a(\s +2\pi )=X^a(\s )$
is given by
\eqn\ssol{
X_m^i(\s ,\tau )= x_m^i +\a'  p_m^i\tau
+i\sqrt{ {\a'  \ov 2} }
\sum_{k\neq 0 }{1\ov k }\bigg(
\a_{(k,m)}^i e^{-i k(\tau -\s )} + 
\tilde \a_{(k,m)}^i e^{-i k(\tau +\s )} \bigg) \ ,
}
\eqn\alco{
[\a_{(k,m)}^i, \a_{(l,n)}^j]=k \delta _{k+l} \delta _{m+n}\delta ^{ij} \ ,
}
$$
\a_{(k,m)}^\dagger=\a_{(-k,-m)}\ ,\ \ \ x_m^\dagger=x_{-m}\ ,
\ \ \ p_m^\dagger= p_{-m} \ .
$$
Throughout, indices $m,n $ are used for  Fourier modes in $\rho $, whereas
$k,l$ are associated with Fourier modes in $\s $.

The wave function of the unperturbed  Hamiltonian $H_0$ is just
the product of an infinite set of harmonic oscillator
wave functions.
The picture of multiple weakly interacting strings is a convenient description
only in the large radius regime.
In sect.  5 we will examine under which conditions
can $H_1$  be treated as a perturbation, and what are 
the relevant degrees of freedom   as the radius is gradually changed.

\subsec{  Instability modes }

Taking $Y_{(0,1)}$ as a Cartan generator,
it follows from the  discussion of sect.~2  that our system 
will have potential valleys along $X_{(0,n)}^a$, corresponding to the
constant modes in the coordinate $\s $.
This holds for any value of the radius $R_0$ (see also sects. 4 and 5).
Let us see what this implies for the solution
 \ssol , valid for large $R_0$.
 Inserting \ssol\ into  eq.~\bbss , we find
\eqn\ooo{
X^a(\s ,\rho, \tau )= x_0^a +\a'  p_0^a\tau +
x^a (\rho ) +2\pi \a'  p^a(\rho )\tau
}
$$
+ \ i\sqrt{ {\a'  \ov 2} }\sum_{k\neq 0 }{1\ov k }\bigg(
\a_k^a(\rho ) e^{-i k(\tau -\s )} + 
\tilde \a_k^a(\rho ) e^{-i k(\tau +\s )} \bigg)\ ,
$$
where 
$$
x^a (\rho )=\sum_{m\neq 0} x_m^a e^{im\rho }\ ,\ \ \ \
p^a (\rho )={1\ov 2\pi }\sum_{m\neq 0} p_m^a e^{ im\rho }\ ,
$$
$$
\a_k^i(\rho )=\sum_{m} \a_{(k,m)}^i e^{im\rho }\ ,\ \ \ \ 
\tilde \a_k^i(\rho )=\sum_{m } \tilde \a_{(k,m)}^i e^{im\rho }\ ,
\ \ \ \alpha^{10}_k=\tilde \a^{10}_k=0\ .
$$
We have separated $x_0^a$ and $p_0^a$, representing the center-of-mass
coordinate and momentum of the membrane,
$$
x_0^a+\a'  p_0^a\tau  ={1\ov 4\pi ^2} \int d\s d\rho X^a(\s,\rho )\ , \ \ \ 
p_0^a=\int d\s d\rho P^a(\s, \rho )\ .
$$
Since the  Cartan subspace is  generated  by $Y_{(0,m)}=e^{im \rho}$,
 the modes that will cause
the instabilities of the supermembrane are  $x^a(\rho)$, $p^a(\rho )$,
i.e.  there are flat directions along all $x_m^a $ with $m\neq 0$.
[By virtue of translational invariance, the zero-mode of $X_0^a(\s ,\t )$ 
does not appear in the Hamiltonian;
the center-of-mass momentum $\int d\s P_0^a$ is exactly conserved, and 
$(p_0^a)^2$ will later be absorbed into the Lorentz-invariant
mass-squared operator.]
 It is important that the $x_m^a$ do not appear in $H_0$.
These are indeed the only directions that are not stabilized
by the winding contributions. It may be less obvious that
 the potential valleys remain in spite of their presence  
in $H_1$, but this can be understood by  closer inspection of
the potential \bbb\ (see  sect. 5).

\newsec{  A simple model }

There are two questions that we  would now like to address.
The first one concerns the regime of applicability of perturbation theory.
The second, in a sense, complementary question is what modes
can be safely dropped for a given radius $R_0$.
Obviously, the modes $x_m^a$ cannot be dropped at any radius,
since they can be excited by an arbitrarily small energy $\epsilon $.
The essential features can be illustrated by the following Hamiltonian:
\eqn\hhaa{
{\cal H}={\cal H}_0 +{\cal H}_1   \ ,
}
\eqn\toyy{
{\cal H}_0= {1\ov 2T ^2}\big( p_{x_1}^2 + p_{x_2}^2 +  p_y^2 \big)+ \ha x_1^2+\ha x_2^2 \ ,
}
\eqn\tty{
{\cal H}_1={\lambda\ov R_0^2}\big( x_1^2 y^2+x_2^2y^2 + x_1^2 x_2^2 \big)
\ ,\ \ \ \l>0\ .
}
The coordinate $y$ is the analogue of the  $x_m^a$, corresponding to the Cartan
generators $Y_{(0,m)}=e^{im\rho }$.
Classically, a mode can  leak out to infinity
along the direction $y$, with $x_1=x_2=0$.
In the quantum theory, the purely bosonic system is stabilized thanks to
a zero-point energy contribution from transverse fluctuations in the
$x_1, x_2$ directions.
In the supersymmetric system, this zero-point energy cancels out;
 as a result, a wave packet can move off to $y=\infty $, and the
spectrum is continuous.
Since these excitations can be produced by an
infinitesimal energy,  the $y$ direction does not get frozen
at any radius $R_0$.
The Hamiltonian \hhaa\ (more precisely, its supersymmetric version) 
 will thus have a continuum spectrum of eigenvalues.

Now  consider the truncated version  
${\cal H}={\cal H}_0 +{\cal H}_1$, with ($T =1$)
\eqn\ttoyy{
{\cal H}_0= \ha ( p_{x_1}^2 + p_{x_2}^2 + x_1^2+x_2^2 ) \ ,\ \ \ \ 
{\cal H}_1={\l \ov R_0^2} x_1^2 x_2^2 \ .\ 
}
In terms of standard creation and annihilation operators, 
\eqn\ttty{
{\cal H}_0=\ha (a_1^\dagger a_1 + a_1 a_1^\dagger +a_2^\dagger a_2 + a_2 a_2^\dagger)\ ,
}
\eqn\annn{
{\cal H}_1={\l \ov 4 R_0^2}(a_1+a_1^\dagger)^2(a_2+a_2^\dagger )^2\ .
}
Let us treat  ${\cal H}_1$ as a perturbation, and write 
the eigenvalues of ${\cal H}$ as $E=E^{(0)}+E^{(1)}+...$, with
$E^{(0)}=n_1+n_2+1$. Since ${\cal H}_0$ has degenerate eigenvalues, to find $E^{(1)}$ we need to solve the secular equation. Denote by $m_1, m_2$
the positive integers satisfying $m_1+m_2+1=E^{(0)}=$~fixed.
Then the $E^{(1)}$ are obtained from the equation
\eqn\secul{
|\langle m_1,m_2|{\cal H}_1|m_1',m_2'\rangle -E^{(1)}\delta _{m_1m_1'}
\delta _{m_2 m_2'}|=0\ ,
}
where 
\eqn\mmm{
\langle m_1,m_2|{\cal H}_1|m_1',m_2'\rangle ={\l \ov 4 R_0^2}
\big[ (2m_1+1)\delta _{m_1m_1'}+\sqrt{(m_1+1)(m_1+2)}\delta_{m_1',m_1+2}+
}
$$
\sqrt{m_1(m_1-1)}\delta _{m_1', m_1-2}\big]
\big[ 1\leftrightarrow 2\big]\ .
$$
Perturbation theory is applicable provided $E^{(0)} \gg E^{(1)}$
(more precisely, $\psi ^{(0)}\gg \psi ^{(1)}$).
For low $E^{(0)}$, all $m_1$ and $m_2$ are small, and the 
solutions to \secul\ will be $E^{(1)}=O(\l /R_0^2 )$.
Thus for those states perturbation theory can be applied
provided $\l/R_0^2 \ll 1$. 
For high $E^{(0)}$,  there will be matrix elements in \mmm \ where both $m_1, m_2\gg 1$, so that  the eigenvalues $E^{(1)}$ can be as high as 
$O(m_1m_2\l/R_0^2)$, with a maximum for $m_1\cong m_2$.
Since $E^{(0)}\cong 2m_1 $, the use of perturbation theory is therefore
justified  only for $ m_1 \ll R_0^2/\l $.
The oscillation amplitude for  states with $m_1, m_2 \gg R_0^2/\l $ is large and
their motion is controlled by ${\cal H}_1$.

As the radius is decreased, the number of modes subject to harmonic motion
will diminish.
For $\l /R_0^2  >1$, even the zero-point mode $n_1=n_2=0$ 
will be influenced by ${\cal H}_1$.
The general behavior of the system  
can be understood by drawing equipotential lines and
 varying $\l/R_0^2$.
For a given energy $E$, the classically accessible region is delimited
by the curve
$$
\ha (x^2_1+x_2^2)+{\l\ov R_0^2} x_1^2x_2^2=E\ .
$$
At $E\ll R_0^2/\l $, 
this line is essentially a circle; the motion is harmonic,
with an oscillation amplitude  given by
$\langle x_1^2 \rangle ,\ \langle x_2^2 \rangle \cong E$.
As $\l/R_0^2 $ is increased, the circles are deformed, with a maximum
decrease of the diameters on the axes $x_1=\pm x_2$. For $\l/R_0^2 \to \infty $,
the figure is extremely narrow, and the motion will be frozen, except
along the axes $x_1=0$, $x_2=0$.
It is easy to see that the only possible finite 
energy excitations will then be those in which the motion takes
 place only in one coordinate, $x_1$ or $x_2$,
that is with $\langle x_1^2 x_2^2\rangle \cong 0$. Indeed, a
wave function $\psi $ where  
$\langle \psi| x_1^2 x_2^2|\psi \rangle $ is non-vanishing will have 
an infinite energy, since
$$
E=\langle \psi| {\cal H}|\psi \rangle >
{\l \ov R_0^2} \langle \psi| x_1^2 x_2^2|\psi \rangle
\to \infty \ \ \ {\rm as}\ \ \ {\l\ov R_0^2}\to \infty \ .
$$
Thus  at very small radius the system reduces to  a 
one-dimensional harmonic oscillator. The only states  relevant to 
low-energy ($E=O(1)$) physics have the form 
$|\psi \rangle =\delta (x_1) |\psi^{(0)}(x_2)\rangle $
or 
$|\psi \rangle =\delta (x_2) |\psi^{(0)}(x_1)\rangle $.


\newsec{  Membrane dynamics at different compactification radii }

The previous discussion can be generalized to our system, with $H_0$ and $H_1$ given by \aaa , \bbb .
As discussed in sect. 3, the  Hamiltonian has a continuum spectrum of eigenvalues due to the zero modes of the strings with $m\neq 0$.
The constant parts of the $X_m^a(\s )$ play the same role as the coordinate $y$ of the model \hhaa : they only appear in $H_1$, but their appearance is not sufficient to remove flat directions.

Let us expand $X_m^a (\s ),\ P_m^a(\s )$ in terms of adimensional
mode operators
\eqn\adim{
 X_m^a (\s )=\sqrt{\a'  } \sum _{k} X_{(k,m)}^a e^{i k \s }\ ,\ \ \ 
 P_m^a(\s )={1\ov 2\pi \sqrt{\a' } } \sum _{k\neq 0} 
P^a_{(k,m)} e^{ ik \s }\ ,\ \ \ 
}
$$
X_{(k,m)}^{10}= P_{(k,m)}^{10}=0\ \  {\rm if }  \ \ k\neq 0\ , 
$$
$$
[ X_{(k,m)}^a, P^b_{(k',m')}]=i\delta_{m+m'} \delta_{k+k'}\delta^{ab}\ ,
$$
where $\a' $ has been introduced in eq. \sredd , and it is related to the
membrane tension by $\a' =(4\pi^2 R_0T_3)^{-1}$.
In the free case, these mode operators will be given as linear combinations
of $\a _{(k,m)}, \tilde \a _{(k,m)}$ operators, with simple exponential
time dependence, as in eq. \ssol .
By inserting the expansions  \adim\ in eqs. \aaa\ and \bbb\ we obtain
\eqn\enmod{
\a'  H_0=\ha \sum _{\n} \big[  P^a_{\n}  P^a_{-\n}
+k^2 X^i_{\n}  X^i_{-\n} \big]\ ,\ \ \ \ \ \ \ \n\equiv (k,m)\ ,
}
\eqn\nome{
\a'  H_1= {1\ov 4g^2}\sum_{\n_1, \n_2,\n_3} 
(\n_1 \times \n_2)(\n_3\times \n_4)
  X_{\n_1}^a  X_{\n_2}^b 
  X_{\n_3}^a   X_{\n_4}^b  
}
$$
+{i\ov g} \sum_{k,m,n} m k^2 X^{10}_{(0,m)} X^i_{(k,n)} X^i_{(-k,-m-n)}\ ,
$$
$$
g^2\equiv {R_0^2\ov \a' }=4\pi^2R_0^3T_3\ ,\ \ \ \ \ \n_4=-\n_1-\n_2-\n_3
\ . 
$$
In the infinite radius limit, $g^2\to \infty $, and the Hamiltonian
becomes that of an  infinite set of harmonic oscillators labelled by
$(k,m)$. It is naturally organized
as an infinite sum of free string theory Hamiltonians labelled by $m$,
which is already manifiest in eq.~\aaa .
Strings with $m\neq 0$ are
the analogue of Kaluza-Klein modes, which  decouple from   low-energy physics
at small compactification radius.

Now the Hamiltonian is  more complicated
than that of the simplified model \hhaa , since it contains an infinite
number of terms and the structure (dictated by the group of
area-preserving diffeomorphisms) is somewhat intricate.
Nevertheless, it is possible to determine under what conditions
a generic quantum state will be weakly coupled for a given (large) $g^2$.
In terms of creation/annihilation operators, $H_0$, $H_1$ contain
terms of the form
\eqn\anni{
\a'  H_0=k N_{(k, m)} + ...\ , \ \  \ 
\a'  H_1={1\ov 4g^2} {(nl-mk)^2\ov kl } N_{(k, n)}N_{(l, m)} +... \ ,
 }
where $N_{(k, m)}=a^\dagger_{(k,m)} \cdot a_{(k, m)}$ is the occupation number
of the oscillator ${(k,m)}$, and we are considering particular terms in $H_1$ with $l'=-k$ and $p=-n$.
Let us take 
$$
n=m_0 \bar n \ , \ \ \ m=m_0\bar m \ ,\ \ \ \ \  
N_{(k,n)}=N_0 \bar N_{(k,n)}\ ,\ \ \  N_{(l, m)}=N_0 \bar N_{(l, m)}\ ,
$$
where $\bar n, \bar m, \bar N_{(k,n)} , \bar N_{(l, m)}$ are $O(1)$ and
$m_0\gg 1$, $N_0\gg 1$. Let us  also consider $l\sim k\gg 1$ and $g^2\gg 1$.
Such a quantum state is weakly coupled provided
$H_1\ll H_0$, or
\eqn\ggmm{
{1\ov g^2} m_0^2 N_0 \ll k \ .
}
It is interesting to note that, for a given string $m_0$ and 
a given occupation number $N_0$, quantum states made of oscillators 
with higher frequency $k$ will describe a more harmonic motion.
This is not a surprise, since the oscillation amplitudes
go like $\langle X_{(k,m)}^2\rangle \sim  N_0/{k} $,
so the region of non-linearities is not reached for high enough $k$.
States with low occupation number made of $a^\dagger _{(k,m)}$,
with $m=O(1)$,  will
be governed by the harmonic oscillator Hamiltonian for all $k$ 
provided $g^2\gg 1$.
From eq. \ggmm\ one also sees that  for any given state $|\psi \rangle $, 
there exists a sufficiently large coupling $g^2$
so that $\langle \psi |H_1|\psi \rangle $ can be neglected
as compared to $\langle \psi |H_0|\psi \rangle $.

It is also possible to estimate the number of strings
that are weakly coupled up to a given string level $N_m=\sum_k k N_{(k,m)}$.
For this it is sufficient to look at states with $k=1$, which have the strongest
coupling for fixed $m$ (i.e.  if the condition \ggmm\ is satisfied for
states made with oscillators of frequency $k=1$, 
then it will be satisfied for
all other states). Then $N_m=N_{(1,m)}= N_0\bar N_{(1,m)}$, so that
eq. \ggmm\ becomes $m^2N_0\ll g^2 $. Thus the number of strings $N_s$
whose excitations below a given level are weakly coupled is given by
$N_s=O(g/ \sqrt{N_0} )$.
Similarly, for a given string $m$,
all quantum string states with 
$$
\a' M^2 \cong N_m \ll O\bigg( {g^2\ov m^2} \bigg)\ ,
$$
will be weakly coupled. States with 
$\a' M^2 > O\big( {g^2/ m^2}\big) $ and $k=O(1)$ will not be subject to harmonic motion,
 since the motion will be mostly governed
by $H_1$.

Let us now see what states  survive in the limit $g^2\to 0$.
By construction, $ H_1$ is a sum of positive-definite
terms; any state $|\Psi \rangle $ with 
$\langle \Psi | H_1 |\Psi \rangle \neq 0$ will have infinite
mass in the zero-coupling limit.
In the limit $g^2=0=R_0^2$, with $T_3\to \infty $ so that 
$T =2\pi R_0 T_3$ remains fixed,
the only states which survive  are
those containing excitations in a Cartan subspace of the area-preserving
diffeomorphism algebra, so that $(\n_1 \times \n_2)(\n_3\times \n_4)=0$
and $H_1$ gives no contribution.
The $X_{(k,0)}$ generates a Cartan subspace, implying that
the type IIA superstring survives.
In addition, just as in the toy model, there are excitations
in other  directions (having $\langle \Psi | H_1 |\Psi \rangle =0$)  which  also remain.
They include, in particular, the states constructed with  the  $X_{(0,m)}^a$ modes,
the constant parts of the  $X_m^a(\s )$ strings,
which are responsible of the instability of the supermembrane.




\newsec{ Membrane spectrum at infinite radius}

The eigenvalues of the mass operator are given by
$$
 M^2= 2 p_0^+p_0^- -(p_0^a)^2 =2 H_0 + 2H_1 -(p_0^a)^2\ .
$$
As is usual in Kaluza-Klein theories,  
$p^{10}_0$ has discrete eigenvalues $p^{10}_0= Q/R_0$, $Q\in {\bf Z}$,
which from the ten-dimensional viewpoint are seen as Ramond-Ramond 
charges.
The ten-dimensional mass operator is 
\eqn\masdd{
M_{10D}^2=2 p_0^+p_0^- -(p_0^i)^2 =M^2+ (p_0^{10})^2 \ .
}

Let us now take the limit $R_0 \to \infty $, $T_3\to 0$, with
$T =2\pi R_0 T_3$ fixed. In this limit, $g^2\to \infty $,
so  the  term $ H_1$ can be dropped,
and the fields can be expanded as in eq. \ssol .
The (bosonic part of the) mass spectrum takes the form
\eqn\masu{
\ha \a'   M^2= \ha\a'  \sum _{m\neq 0} p_m^a p_{-m}^a
+ \sum_{m=-\infty }^\infty \sum_{k=1} ^\infty
\big[  \alpha _{(-k,-m)}^i  \alpha _{(k,m)}^i+
\tilde \alpha _{(-k,-m)}^i  \tilde \alpha _{(k,m)}^i \big] \ .
}
In this limit, the fact that the 
standard membrane spectrum is continuous is simply  understood:
the $p_m^i=\int d\s P_m^i(\s )$ (center-of-mass momenta of $m\neq 0$ ``strings")
 take continuous values, since
they are  governed by the free
particle Hamiltonian 
$H_{\rm free}=\ha p_m^i p_{-m}^i =\ha \big( 
p_m^{(1)}\big) ^2+\ha \big( p_m^{(2)}\big) ^2$,
where $p_{\pm m}=p^{(1)}_m\pm i p^{(1)}_m$.

It is  interesting to discuss the spectrum of a truncated 
Hamiltonian, defined by $\aaa ,\bbb $ by setting 
 the
constant parts in $\s $ of the fields $X^a_m(\s ), P^a_m (\s )$ to zero.
Now no flat direction remains in the potential,
and  the motion of all modes is bounded.
The mass spectrum is  discrete for all radii
and, in the
infinite radius limit, takes the simple form 
\eqn\mass{
\ha \a'  M^2= {\bf N}+ {\bf \tilde  N}\ , 
}
where the bosonic part of ${\bf N}, \  {\bf \tilde  N}$ can be read
from eq. \masu .
 
Let us now restore the fermion contributions in the formulas.
Inserting the expansion
\eqn\ttdd{
\sqrt{p_0^+}
\theta ^\a (\s ,\rho )=  \sum _m S_m^\a (\s ) e^{im\rho }\ ,
}
in $H_{\rm F}$, we get (see eq. \lcgh )
\eqn\fffk{
H_{\rm F}= {T }\int d\s \bigg[ \sum_m  
\bar S_{-m}\Gamma^{10} \del_\s S_m
-\sum_{m,n} {i\ov R_0} \bar S_{-m-n}\Gamma^{a}\big( n \del_\s X^a_m S_n-
 m X^a_m \del_\s S_n \big) \bigg] \ .
}
As $R_0\to \infty $, the second term disappears and the Hamitonian
reduces to that of a free system.
The ten-dimensional  chiralities of the fermions are those of
type IIA superstring theory.
The complete mass formula is thus given by eq.~\mass , with  
${\bf N},  {\bf \tilde  N} $ containing the standard
Green-Schwarz fermion-mode operators $S_{(k,m)}^r, \td S_{(k,m)}^r$,
$r=1,...,8$,
for the different strings parametrized by $m$, that is
\eqn\nntn{
{\bf N}=\sum_{m=-\infty }^\infty \sum_{k=1} ^\infty
\big[ \alpha _{(-k,-m)} ^i  \alpha _{(k,m)} ^i +
k S _{(-k,-m)}^r  S _{(k,m)}^r \big] \ ,
}
$$
\{ S^r _{(k,m)}, S^s _{(k',m')} \} =\delta_{m+m'}\delta_{k+k'}\delta^{rs }
\ ,
$$
and a similar  expression for ${\bf\tilde  N}$. The zero-point energy
cancels out as in the usual Green-Schwarz light-cone formalism
of the superstring.

The physical Hilbert space is spanned by states made of the 
transverse excitations
$\alpha _{(-k,m)} ^i, \ \td \alpha _{(-k,m)} ^i $ and 
$\ S _{(-k,m)} ^r,\ \td S _{(-k,m)} ^i$, with $k>0$ and  $m\in {\bf Z}$.
In addition,  physical states must obey the level matching
conditions:
\eqn\fffs{
{\bf N}={\bf \td N}\ ,\ \ \ \ \ {\bf N^+}-{\bf  N^-}=Q\ .
}
The definition of the operators ${\bf N^+}, {\bf  N^-}$ is given in the appendix, where  the derivation of conditions \fffs\ is given.
The first of these conditions plays the same role as the standard
Virasoro level-matching condition of superstring theory.
The second condition is new and it has no analogue in superstring theory. 
It is associated with translational invariance in the $\rho $ direction.
Note that physical states with Ramond-Ramond charge $Q\neq 0$ necessarily 
involve excitations in
the $\rho $ direction, i.e. they have ${\bf N^+}\neq 0$ or ${\bf  N^-}\neq 0$.

It may seem surprising that in the infinite radius limit the membrane spectrum takes the simple form \mass , with the 
constraints \fffs .
However, it should be stressed that this is the spectrum for a
membrane that wraps around the compact dimension, in the specific limit
$T_3\to 0 $, $R_0\to \infty $, with $R_0T_3 $ fixed. In this limit 
the quartic terms 
in the potential are negligible in relation to the quadratic terms.
For a membrane that does not wrap around $S^1$, there is no quadratic term
and at any radius the dynamics is governed by the quartic terms.
Note also that this membrane configuration does not give extra contributions
to the energy. 
For a target space \rss , it is possible to have the topologically non-trivial membrane winding number  \wiwi  , which gives a contribution to the mass squared
proportional to $T_3^2 w_0^2 R_9^2 R_{10}^2$, as in eq.~\aaaa .
 It is of interest to study the spectrum in the case \rss ,
since it will allow a comparison between the BPS excitation spectrum 
of type IIB superstring theory on ${\bf R}^9\times S^1$ 
and the M-theory counterpart,  given in terms
of wrapped 2-branes.\foot{
This calculation has been recently carried out in \rutse , and
an exact matching of the corresponding BPS spectra was found.}
The correspondence between Kaluza-Klein modes and
winding states of type IIB  and M-theory was investigated by Schwarz 
(see \schw\ and additional references there).
 

\newsec{ Consequences for M-theory}

What has been done in the previous sections may also be interpreted
as follows.
Suppose we start from type IIA superstring theory, imagining that this should
arise
as the zero-radius limit of some eleven-dimensional theory on
\rs . We then turn on a small radius $R_0=\epsilon $, and inquire what 
kind of new degrees of freedom could be consistently added.
The double-dimensional reduction  concretely 
indicates that we should consider a 2-brane with one leg
wrapped around the compact direction.
However, we have seen that there are states
made of mode operators associated with Cartan directions
(for which $\langle \Psi | H_1 |\Psi \rangle = 0$, see eq. \nome ),
which can never emerge at $R_0=\epsilon $ if they were not already
present at $R_0=0$.
In particular, wave packets constructed with 
the zero modes $X_{(0,m)}^i, \ P_{(0,m)}^i$. 
These zero modes are  part of the fundamental scalar fields of the 
(2+1)-dimensional theory, and they cannot be  excluded from the Hamiltonian, unless
there is a  kinematical or dynamical  reason for this.\foot{
Interestingly, in open string theory, there is a consistent
way to quantize strings in which the center-of-mass remains fixed.
This is done by imposing Dirichlet boundary conditions
at the endpoints \dlp . As a result, 
the center of mass, 
$x= {1\ov \pi }\int  d\s  X(\s ,\tau )$ 
is  $\tau $ independent. 
Some sort of closed string analogue  (e.g.   
where the $p_m^a= \int d\s P^a_m(\s )$ are forced to be zero)
may be of relevance to the present case.
} 

It may be worth  emphasizing that, unlike the spectrum of superstring theory,  
the spectrum \mass\ 
contains the infinite tower of Kaluza-Klein quantum states 
carrying Ramond-Ramond charges. 
For small radius, the ten-dimensional mass operator \masdd\ will be of
the form
\eqn\sssa{
\ha \a'  M^2_{\rm 10D}= { Q^2\ov 2g^2 }
+ {\bf N}+ {\bf \tilde N}+  \a'  H_1 \ ,\ \ \ \ g^2=R_0^2/\a' \ ,
}
where the $m\neq 0$ part of ${\bf N}, {\bf \tilde N}$ will be expressed
in terms of $X^i_{(k,m)}, P^a_{(k,m)}$, as in eq.~\enmod . 
Quantum states with $Q=0$ containing excitations of 
$a^\dagger _{(k,m)}$ with $m\neq 0$ will 
have a mass  of order $1/g$, originating from contributions of $H_1$.
In addition, 
there are  Kaluza-Klein states with $Q\neq 0$,
which have $\sqrt{\a' } M_{\rm 10D}= O(1/ g )$.
Note that they must contain excitations with $m\neq 0$ in order to
satisfy the second constraint in eq.~\fffs .
These states are presumably candidates for D0-branes and
1-branes of the type IIA superstring theory.

It would also be interesting to investigate the zero-mode problem in terms
of  dual actions. In particular, the duality between  D2-brane action 
and the supermembrane action suggests that a more adequate description
for the small radius limit could be in terms of a Born-Infeld type action, where
the variables include world-volume vectors \toww .
In this small $g^2$ limit, one could alternatively 
consider the 5-brane action.
Another possibility is that the extra modes are just confined 
(with energy gaps $M$ of order $1/g$) thanks to $\s $-model quantum corrections, which should
be important for $R_0< (T_3)^{-1/3}$. Unfortunately, little is known about higher-derivative ($\k $-invariant) extensions of the supermembrane
action, but it would certainly be useful to have a tractable example.

\vskip 1cm


\vfill
\eject

\appendix{A}{Global constraints and the physical Hilbert space}

In the light-cone gauge, the coordinate $X^+$ is set to
$X^+=p^+_0\tau $, 
and  $X^-$ is determined from the constraints
in terms of the other cordinates (see e.g. \refs {\bst - \duff }):
\eqn\aauu{
\del_{\bar a } X^-= {1\ov p_0^+} \del_{\bar a } X^a 
\dot X^a\ ,\ \ \ \ \bar a=\s, \rho\ .
}
By taking the curl, one gets the condition
\eqn\xxaa{
\{ X^a, \dot X^a\} =\del_\s X^a \del _\rho \dot X^a -
\del_\rho X^a \del _\s \dot X ^a
\equiv 0\ ,
}
or, in phase-space variables,
\eqn\xdaa{
\{ X^a, P^a\}\equiv 0\ .
}
The Fourier components
\eqn\fouu{
L_\n ={1\ov 4\pi ^2} \int d\s d\rho\ Y^\n \{ X^a, P^a\}\ 
}
are generators of the algebra of area-preserving 
diffeomorphisms \lieb .
By a gauge transformation we can always rotate one of the coordinates,
say $X^{10}$, into the Cartan subspace, as in eq.~\uno .
Then eq.~\xdaa\ takes the form
\eqn\acinco{
L_{\bf m} ^{\rm T}=-\sum_{\n '} \sum _{\n \in K} 
X^{10}_\n P^{10}_{\n '} f_{\n \n ' {\bf m} }\ ,\ \ \ \ \ 
L_{\bf m} ^{\rm T}\equiv {1\ov 4\pi ^2} \int d\s d\rho\ Y^{\bf m} \{ X^i, P^i\} \ .
}
We note that the $P_{\n '}^{10} $, $\n ' \in K$, are absent from this formula.
The constraint \acinco\   determines  
$\td P^{10}(\s ,\rho ) \equiv \sum _{\n \not \in K} P_\n^{10} Y_\n (\s ,\rho )$
in terms of the $X^i , \ P^i $ and $X_K^{10}$.
By formally inverting eq. \acinco , one gets
\eqn\aseis{
\td P^{10}_{\n }=\{ F^{-1}\} _{ {\bf m} \n } L_{\bf m} ^{\rm T}\ , \ \ \ \ 
F _{\n{\bf m }}= - \sum _{\n ' \in K} X^{10}_{\n'} f_{\n ' \n  {\bf m} }\ .
}
The determinant of $F$ vanishes when some of the eigenvalues of $X^{10}_{\n'} $
coincide, i.e. 
at the boundary of the Weyl chamber \dewit .
In the present case of a membrane wrapped on \rs , 
$$
X^{10}=R_0\rho + \tilde X^{10} ( \rho )\ ,
$$
and for small oscillation amplitudes, 
$\langle |\del_\rho \tilde X^{10}|\rangle \ll R_0$, 
the relation \acinco\ is always invertible,  and
a simple solution
of the constraint eq.~\xxaa\ 
can be found.  
Equation \xxaa\ becomes
\eqn\xxbb{
\{ X^i, P^i\} = \big (R_0+ \del_\rho  \tilde X^{10}  \big) \del_\s   \tilde  P^{10} 
}
In the $R_0\to \infty $ limit (with fixed winding around the compact direction), 
this  reduces to
\eqn\xxcc{
\{ X^i, P^i\} \cong   R_0 \del_\s \td  P^{10}\  ,
}
which determines $\td  P^{10}$ in terms of transverse coordinates.
In the opposite limit $R_0\to 0$, there are singular boundary points which
need to be treated
with care. The discussion of instability modes
is nevertheless not affected, since one can always choose suitable
wave packets with support in 
the interior of the Weyl chamber \dewit .

Since $X^-$ is single-valued, eq. \aauu\ also implies the global constraints
\eqn\vvs{
{\bf P^{(\s )} }={1\ov 2\pi \a' } 
\int _0^{2\pi } d\s \ \del_\s X^a \dot X^a \equiv 0\ ,
}
\eqn\vvr{
{\bf P^{(\rho )} }= {1\ov 2\pi \a' }
\int _0^{2\pi } d\rho \ \del_\rho X^a \dot X^a \equiv 0\ .
}
The operators $ {\bf P^{(\s )} },\ {\bf P^{(\rho )} } $ generate
translations in $\s $ and $\rho $, respectively.
By virtue of eq. \xxaa , the integrals in \vvs , \vvr\ are independent of the contours.
In particular, one readily checks that
\eqn\taaa{
\del_\rho {\bf P^{(\s )} }=0\ ,\ \ \ \ \ \ \ \del_\s {\bf P^{(\rho )} }=0\ .
}
Let us now compute ${\bf P^{(\s )} }$ and ${\bf P^{(\rho )} }$ to leading order  in the large-radius limit. By making use of the properties \taaa , we can write
${\bf P^{(\s )} } ,\ {\bf P^{(\rho )} }$ in the more convenient form:
\eqn\fff{
{\bf P ^{(\s )} } 
={1\ov 4\pi ^2\a'  }\int _0^{2\pi } d\rho \int _0^{2\pi } d\s \  \del_\s X^a \dot X^a \ ,\ 
}
\eqn\fgg{
{\bf P^{(\rho )} }={1\ov 4\pi ^2\a'  }\int _0^{2\pi } d\s \int _0^{2\pi }
d\rho \ \del_\rho X^a \dot X^a \ .\ 
}
Inserting eqs. \bbss, \ssol\  in eqs. \fff\ and \fgg , 
after some straightforward algebra,
we find the following expressions:
\eqn\ccvv{
{\bf P ^{(\s )} }= 
- \sum_{m=-\infty }^\infty \sum_{k=1} ^\infty
\big[ \alpha _{(-k,-m)} ^i  \alpha _{(k,m)} ^i 
- \td \alpha _{(-k,-m)} ^i  \td \alpha _{(k,m)} ^i \big] \ ,
}
\eqn\ccbb{
{\bf P ^{(\rho )} }=Q+ {i} \sum_{m=1 }^\infty m
\big( x_m^i p_{-m}^i -x_{-m}^i p_{m}^i\big)
-
\sum_{m,k=1} ^\infty
{m\ov k} \big[ 
\alpha _{(-k,-m)} ^i  \alpha _{(k,m)} ^i   
}
$$
+\td \alpha _{(-k,-m)} ^i  \td \alpha _{(k,m)} ^i 
-\alpha _{(-k,m)} ^i  \alpha _{(k,-m)} ^i  
-\td \alpha _{(-k,m)} ^i  \td \alpha _{(k,-m)} ^i 
\big] \ ,
$$
where we have used 
$$
X^{10}= R_0 \rho + \sum_m \big(x^{10}_m+\a' p_m^{10}\tau \big) e^{im\rho }
\ , \ \ \ \ p^{10}_0=Q/R_0\ .
$$
It is convenient to introduce mode operators
$$
\beta _{(k,m)}^i={1\ov \sqrt{|k|} }\alpha _{(k,m)}^i\ ,
\ \ \ \ \td \beta _{(k,m)}^i={1\ov \sqrt{|k|} }\td \alpha _{(k,m)}^i\ ,
$$
so that the commutators \alco\ and eqs. \ccvv , \ccbb\ take a more symmetrical form under $k\leftrightarrow m$. 
Restoring the fermion contributions, the  constraints \vvs , \vvr\ 
become
\eqn\lvmv{
{\bf N}={\bf \td N}\ ,
}
\eqn\masp{
{\bf N^+}-{\bf  N^-}= Q \ ,
}
where
\eqn\zzxx{
{\bf N}=\sum_{m=-\infty }^\infty \sum_{k=1} ^\infty
k \big[  \beta _{(-k,-m)} ^i  \beta _{(k,m)} ^i + 
 S _{(-k,-m)}^r  S _{(k,m)}^r \big]\ ,
}
\eqn\tdtdd{
{\bf \td N}=\sum_{m=-\infty }^\infty \sum_{k=1} ^\infty
k \big[  \td \beta _{(-k,-m)} ^i  \td \beta _{(k,m)} ^i +
 \td S _{(-k,-m)}^r  \td S _{(k,m)}^r \big]\ ,
}
\eqn\zzmm{
{\bf N^+}={\bf N_0}+ \sum_{m=1 }^\infty \sum_{k=1} ^\infty 
m \big[  \beta _{(-k,-m)} ^i  \beta _{(k,m)} ^i 
+  S _{(-k,-m)}^r  S _{(k,m)}^r 
}
$$
+ \td \beta _{(-k,-m)} ^i  \td 
\beta _{(k,m)} ^i + \td S _{(-k,-m)}^r  \td S _{(k,m)}^r
\big]\ ,
$$
\eqn\zzmp{
{\bf N^-}= - {\bf N_0 } + \sum_{m=1 }^\infty\sum_{k=1} ^\infty 
m \big[  \beta _{(-k,m)} ^i  \beta _{(k,-m)} ^i 
+ S _{(-k,m)}^r  S _{(k,-m)}^r 
}
$$
+ \td \beta _{(-k,m)} ^i  \td \beta _{(k,-m)} ^i 
+  \td S _{(-k,m)}^r  \td S _{(k,-m)}^r 
\big]\ ,
$$
$$
{\bf N_0}=
{1\ov 2} 
\sum_{m=1 }^\infty  m \bigg[
i \big( x_{-m}^a p_{m}^a -x_{m}^a p_{-m}^a\big)
+ \big[ S_{(0,m)}^r, S_{(0,-m)}^r \big]
+ \big[ \td S_{(0,m)}^r, \td S_{(0,-m)}^r \big] \bigg] \ .
$$
In the truncated theory,  zero modes of $m\neq 0$ strings are dropped
and  one can set ${\bf N_0}=0$.

The vacuum Fock state is defined by
\eqn\focks{
\alpha _{(k,m)}^i|0\rangle =\td \alpha _{(k,m)}^i|0\rangle = 0\ ,\ \ \ \ k>0\ ,
}
\eqn\sssa{
S _{(k,m)}^r |0\rangle =\td S _{(k,m)}^r |0\rangle = 0 \ ,\ \ \ \ k>0\ ,
}
$$
p_m^a|0\rangle =0\ ,
$$
for all $m$.
The Fock space is generated by the states made by successive applications of
the $\alpha _{(-k,m)}^i,\ S _{(-k,m)}^r, \ \td \alpha _{(-k,m)}^i,\ 
\td S _{(-k,m)}^r$,  $k>0$ on the vacuum Fock state.
Thus, in the $R_0\to \infty $ limit, we find that the Hilbert
physical space is constituted of all
states in the Fock space  obeying the conditions \lvmv \ and  \masp .

\listrefs

\bye